# Assessing and Improving Machine Learning Model Predictions of Polymer Glass Transition Temperatures

Manav Ramprasad[1] and Chiho Kim[2]
[1]Joseph Wheeler High School, 375 Holt Road, Marietta, GA 30068, US
[2]School of Materials Science and Engineering, Georgia Institute of Technology, 771 Ferst Dr. NW, Atlanta, GA 30332, US

The success of the Materials Genome Initiative has led to opportunities for data-driven approaches for materials discovery. The recent development of Polymer Genome (PG), which is a machine learning (ML) based data-driven informatics platform for polymer property prediction, has significantly increased the efficiency of polymer design. Nevertheless, continuous expansion of the 'training data' is necessary to improve the robustness, versatility and accuracy of the ML predictions. In order to test the performance and transferability of the predictive models presently available in PG (which were previously trained on a dataset of 450 polymers), we have carefully collected additional experimental glass transition temperature ($T_g$) data for 871 polymers from multiple data sources. The $T_g$ values predicted by the present PG models for the polymers in the newly collected dataset were compared directly with the experimental $T_g$ to estimate the accuracy of the present model. Using the full dataset of 1321 polymers, a new ML model for $T_g$ was built following past work. The RMSE of prediction for the extended dataset, when compared to the earlier one, decreased to 27 K from 57 K. To further improve the performance of the $T_g$ prediction model, we are continuing to accumulate new data and exploring new ML approaches.

## Introduction

Polymers, displaying a dizzying diversity of physical and chemical properties, constitute an important and ubiquitous class of materials (1). Although they are made up of a certain number of atomic species found from the periodic table, such as C, H, and O, this seemingly limited chemical palette leads to a rich and diverse spectrum of distinct polymers with a broad range of property values. Thus, it is highly non-trivial to find a suitable optimal polymer for a particular application with desired properties in the practically infinite chemical space. As a result, selection of polymers has hitherto proceeded largely by intuition and trial-and-error efforts, which generally tend to shape the materials discovery landscape in a painstakingly slow manner.

In 2011, the White House unveiled the Materials Genome Initiative (MGI) to accelerate the discovery, manufacture and deployment of advanced materials twice as fast as in the past but at a fraction of the cost (2). One of the central pillars of the MGI is to utilize data-driven approaches, such as machine learning (ML), to speed up materials discovery, including in polymer science and engineering. Data-driven ML approaches are complementary to traditional approaches practiced in materials science and engineering (3). ML approaches utilize prior data, information and knowledge in an effective and efficient manner, as has been demonstrated in many other domains in the past. Classic examples of the ML approaches include facial, fingerprint or object recognition systems, machines that can play sophisticated games such as chess, Go or poker, and automation systems such as in robotics or self-driving cars (4).

Within the domain of materials science and engineering, the synthesis and testing process in the laboratory tends to be expensive and time-consuming especially when handling the polymeric system. In order to utilize the data-driven framework, a dataset of several similar materials and their properties must be first collected. This data constitutes "prior knowledge" on this situation, i.e., the data is obtained from previously performed dedicated experiments or from the literature. Each of the materials in the dataset is then converted to a unique numerical representation, typically referred to as the "fingerprint". Finally, a mapping is established between the fingerprint and their properties using ML algorithms such as Gaussian process regression (GPR), thus leading to a predictive surrogate model (5). Subsequently, this model can be used to make instantaneous predictions of the properties of a new material, by simply following the fingerprinting and mapping procedures. The essential elements of this workflow are portrayed in **Figure 1**.

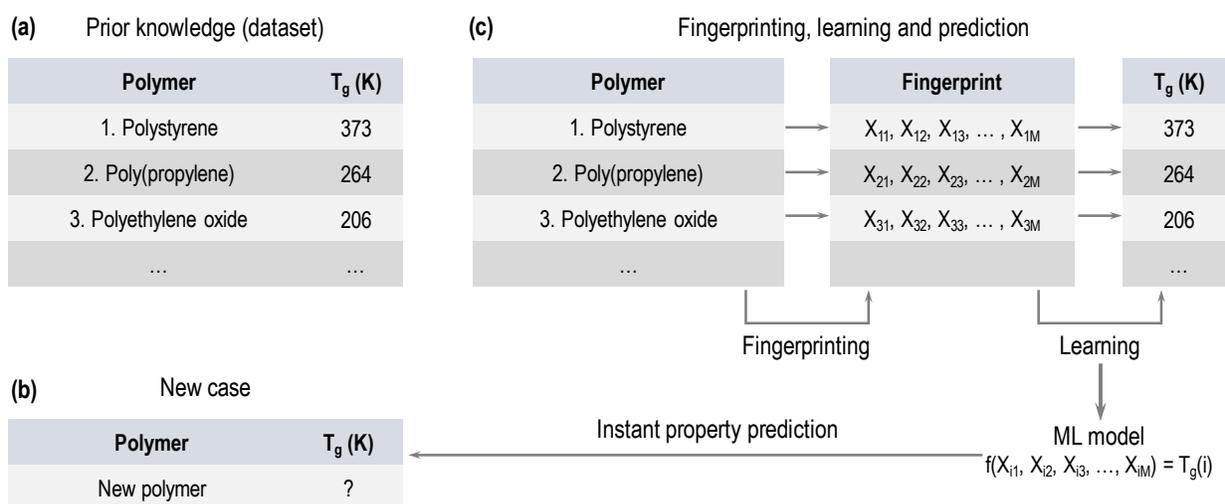

**Figure 1.** The key elements of machine learning in materials science. **a.** Schematic view of an example data set. **b.** Statement of the problem "What is the $T_g$ of new polymer?" **c.** Creation of a prediction model via the fingerprinting and learning steps.

The efficacy of this philosophy has been recently demonstrated as part of the "Polymer Genome" (PG) Project (6). In order to improve upon the predictive capabilities of the ML models implemented, data collection is extremely important. The present work deals with testing the capability of PG on new polymers, and using the results of this test to improve the predictive models. The property chosen for this test is the glass transition temperature, $T_g$, the temperature above which a polymer transitions from glass-like to rubber-like. $T_g$ is an enormously important property in many applications, as it determines the temperature ranges at which it is safe to use a polymer. Previously, the model hosted by PG was trained on 450 polymers. Current work demonstrates how the expansion of the dataset affects the performance of the ML model. We have collected additional experimental $T_g$ data for 871 polymers. The predictions of PG for these new polymers were compared directly with the collected $T_g$ data, and conclusions have been drawn regarding the deficiencies of PG. The original training set was then augmented with this new data, retraining was performed, and this has led to an improvement in the predictive capability of PG.

**Results**

Although efficient, ML models are accurate and reliable only within the domain of the dataset the model was trained on. Predictions made for cases that fall outside the domain of the training data (i.e., the dataset originally used to create the models) are not expected to be reliable. In such cases, the new data points that fall outside the original domain of applicability have to be necessarily included in a retraining process to make the predictive model more versatile and transferable.

We refer to the earlier version of PG that was trained on 450 $T_g$ values as PG-0. The newer version of PG in which the new $T_g$ data for 871 additional polymers has been infused is referred to as PG-1 (Details of data distribution and example polymers in the dataset are shown in the section **Methods**). Since PG-0 was trained on the original 450 data points, the predictions for those 450 points are fairly accurate. The prediction for the new polymers, on the other hand, is inaccurate and uncertainty of the prediction is higher. **Figure 2(a)** shows a parity plot of the performance of PG-0 on both the new dataset of 871 polymers and the initial 450. As can be seen, while many polymers fall on the parity line indicating good agreement between predicted and actual values, predictions for a certain portion of new polymers are off the parity line.

The poor predictive capabilities for those points in the range 300 K - 500 K is mainly due to the difference of fingerprint for the new data points compared to that of the benchmark data points. In the case of very high $T_g$ values, the PG-0 model performs poorly due to a lack of benchmark data points in the high $T_g$ region (see also Figure 3 that shows the distribution of $T_g$ values found in the original and the new datasets). In all cases for which the predictions are poor, the uncertainty of the predictions, which is depicted by error bar around data points in **Figure 2**, is relatively higher than those for the original 450 polymers. Higher uncertainty for a particular case indicates that the polymer is 'not very similar' to the 450 training set polymers of PG-0. Overall, the performance in terms of the root mean square error (RMSE) for PG-0 is greater than 50 K for the set of new 871 polymers. This RMSE is higher than desired for $T_g$ predictions (below 30 K). This observation indicates that more data points are necessary to improve the predictive performance of the ML model.

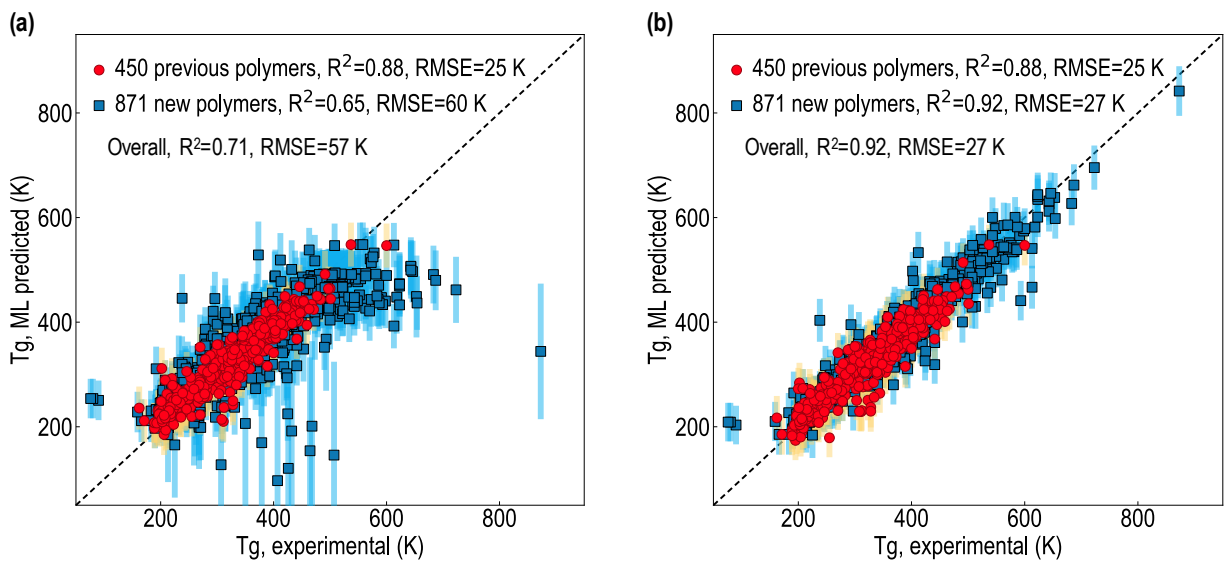

**Figure 2.** Performance of ML prediction model. Comparison of models trained on (a) 450 previous polymers and (b) 1321 polymers including 871 new polymers. Error bar represents GPR uncertainty (confidence of prediction).

Next, the 871 new polymers and their corresponding $T_g$ values were used to augment the original $T_g$ dataset used for PG-0, followed by retraining to create a new PG-1 GPR predictive model for $T_g$. **Figure 2(b)** shows the performance of the PG-1 model. As can be seen, a remarkable improvement in predictions emerges. The RMSE in this case is well below 30 K which is acceptable, as the uncertainties in the actual measurement of $T_g$ is in the same range. In addition, the uncertainties calculated by GPR shown by the yellow error bars have also decreased significantly, again showing an improvement in prediction capabilities. Relative to the original dataset, the new dataset has specifically added polymers in new chemical spaces, and has added polymers with high $T_g$ values, i.e., in the 500-700K range. These aspects have led to a significantly better predictive capability of PG. Further progress can occur by systematically adding more diverse data.

**Discussion**

Many other properties of polymers, besides $T_g$, are important as well. In addition to the $T_g$ prediction, PG also offers predictions of other properties including 1) electronic properties like band gap, ionization energy and electron affinity, 2) dielectric and optical properties such as the dielectric constant and the refractive index, 3) physical and thermodynamic properties like density and atomization energy, 4) solubility properties like Hildebrand solubility parameter and list of solvent and non-solvent, 5) mechanical properties like tensile strength and Young's modulus, and 6) permeability properties like gas (He, $H_2$, $CO_2$, $N_2$, $O_2$, and $CH_4$) permeability. Each of these predictive models within PG can potentially go through an improvement due to new data infusion.

In summary, to improve upon an existing machine learning model to predict polymer glass transition temperatures ($T_g$), a comprehensive dataset of Polymer $T_g$ was collected. Machine learning predictions for these new polymers revealed the deficiencies of the previous model. In retraining the machine learning model on the new data, the performance of the predictions dramatically improved. This work has thus led to a $T_g$ prediction model that has been exposed to a more diverse dataset than before, and is hence more versatile. The new prediction model presented for $T_g$, as well as the other polymer properties listed above is available for free at the Polymer Genome online platform (https://www.polymergenome.org/). Looking further into the future, it would be useful if the prediction pipeline can be inverted, i.e., if polymers can be recommended that meet a specific set of property objectives, such as $T_g$ between 600 K and 650 K, and band gap between 4 eV and 5 eV. A variety of artificial intelligence based algorithms (7,8) may be utilized for such purposes. Solving this inverse problem effectively will significantly accelerate polymer discovery.

## Methods

Data for this work were obtained from publicly-available collections of experimental measurements (9, 10) and an online repository of polymer properties (11). The polymer dataset is highly diverse and the constituent polymers are composed of nine atomic species: C, H, O, N, S, F, Cl, Br and I. The $T_g$ of the 1321 polymers (450 polymers from previous work and 871 newly collected polymers) in the dataset varied widely, ranging from 76 K to 873 K with a mean of 354 K (**Figure 3**). The repeat unit of the polymers were represented using the simplified molecular-input line-entry system (SMILES) (12). Examples of SMILES representation are shown in **Figure 3(b)** with original name of polymers and experimental $T_g$.

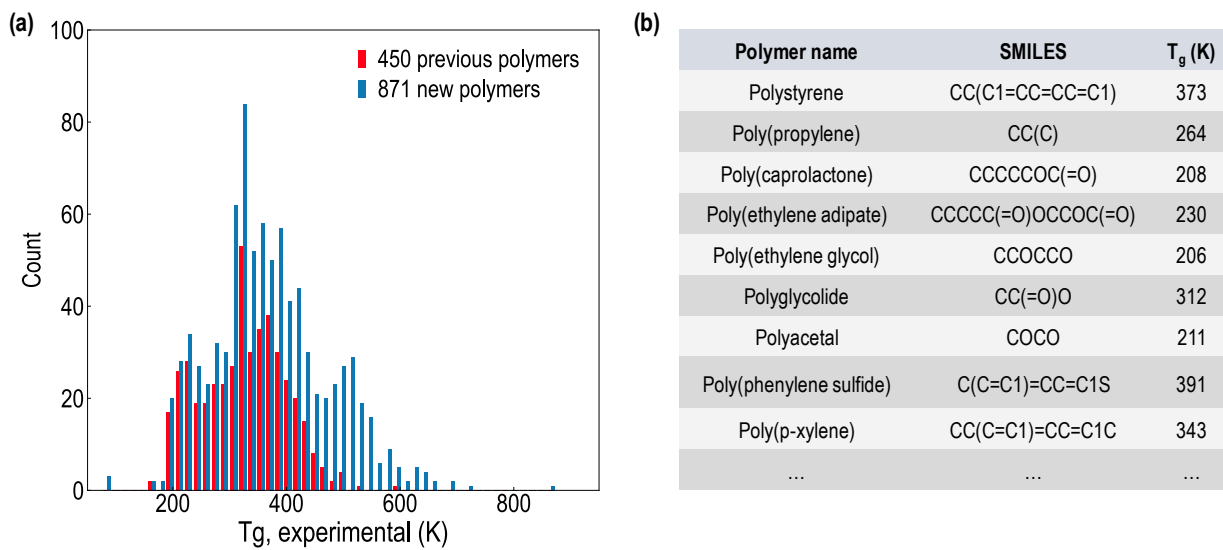

**Figure 3.** (a) Distribution of the $T_g$ values for all the polymers considered in this work. (b) Sample polymer dataset with SMILES representation and experimental $T_g$ values.

In order to capture the key features that may control the $T_g$, we utilized the hierarchical polymer fingerprinting scheme (13). The fingerprint building process consists of three hierarchical levels of features. The first one is at the atomic scale wherein the occurrence of atomic fragments. For 1321 polymers, there are 128 such components. The next level deals with quantitative structure property relationship (QSPR) descriptors (14), such as estimated surface area of polymer repeating unit, and fraction of rotatable bonds. Such descriptors, 39 in total, form the next set of components of our overall fingerprint. The third level descriptors captured morphological features such as the topological distance between aromatic rings and the length of side-chains. We include 22 morphological features in the fingerprint.

The ML model was built by mapping the descriptors to the $T_g$ values using GPR with a sum-kernel consisting of a radial basis function kernel and a white-noise kernel. Within this scheme, data points with fingerprint very close to the new fingerprint value will be weighted more than data points with fingerprints farther away from the fingerprints of the new data point. This means that if the new polymer is similar in terms of fingerprint to some polymers already in the data set used for training the ML model, GPR will give a $T_g$ value close to that of the similar polymers. Details of the approach used may be found in previously published work (13).